\documentclass[reprint, superscriptaddress, amssymb, aps, pra, a4]{revtex4-2}
\usepackage{graphicx}
\usepackage{dcolumn}
\usepackage{hyperref}
\usepackage{braket}
\usepackage{bm}
\usepackage{natbib}
\usepackage{amsmath}
\usepackage{tabularx}
\usepackage{xcolor}
\usepackage{soul}

\begin{document}

\title{Transformation of Bell states using linear optics}

\author{Sarika Mishra} \email{sarika@prl.res.in}
\affiliation{Quantum Science and Technology Laboratory, Physical Research Laboratory, Ahmedabad, India 380009.}
\affiliation{Indian Institute of Technology, Gandhinagar, India 382355.}

\author{R. P. Singh} \email{rpsingh@prl.res.in}
\affiliation{Quantum Science and Technology Laboratory, Physical Research Laboratory, Ahmedabad, India 380009.}

\date{\today}

\begin{abstract}
Bell states form a complete set of four maximally polarization entangled two-qubit quantum state. Being a key ingredient of many quantum applications such as entanglement based quantum key distribution protocols, superdense coding, quantum teleportation, entanglement swapping etc, Bell states have to be prepared and measured. Spontaneous parametric down conversion is the easiest way of preparing Bell states and a desired Bell state can be prepared from any entangled photon pair through single-qubit logic gates. In this paper, we present the generation of complete set of Bell states, only by using unitary transformations of half-wave plate (HWP). The initial entangled state is prepared using a combination of a nonlinear crystal and a beam-splitter (BS) and the rest of the Bell states are created by applying single-qubit logic gates  on the entangled photon pairs using HWPs. Our results can be useful in many quantum applications, especially in superdense coding where control over basis of maximally entangled state is required.
\end{abstract}


\maketitle

\section{\label{sec:intro}Introduction}
\noindent Quantum entanglement \cite{einstein1935can, horodecki2009quantum,dehlinger2002entangled} is one of the most interesting phenomena in quantum world which laid the foundation of many quantum applications such as quantum superdense coding \cite{liu2002general,mattle1996dense}, quantum teleportation \cite{bennett1993teleporting,bouwmeester1997experimental}, entanglement swapping \cite{pan1998experimental,zukowski1993event}, and quantum key distribution (QKD) protocols \cite{mishra2021bbm92,gisin2002quantum,yin2020entanglement,yin2017satellite,kwek2021chip,long2002theoretically}, quantum secret sharing (QSS)\cite{hillery1999quantum, ju2022measurement}, quantum secure direct communication (QSDC)\cite{sheng2022one, ying2022measurement, hu2016experimental,zhang2017quantum} etc. The simplest way to create an entangled photon pair is spontaneous parametric down-conversion (SPDC) process \cite{couteau2018spontaneous, karan2020phase, wang2022manipulation}. In this process, a nonlinear $\chi^{(2)}$ crystal is used to produce two correlated photons. These photon paires can be correlated in any degree of freedom such as, polarization, orbital angular momentum (OAM), energy, time, frequency etc. Among these, polarization entangled states are easy to prepare and measure. Hence it is the most widely used resource for all the quantum communication protocols.\par
A set of all four maximally entangled polarization states in two dimensional Hilbert space is known as Bell states \cite{dehlinger2002entangled}. In many quantum information schemes such as quantum superdense coding, teleportation, and entanglement swapping, the generation and discrimination of all the four maximally entangled states is required. For example, in superdense coding control over basis of maximally entangled state is required. A desired Bell state can be prepared through unitary transformations on an entangled pair of photons generated in SPDC. In this case, transformation of one Bell state into another by using single-qubit gate (Pauli gate) can be used. For example, $\ket{\Psi}^{+}$ can be transformed into $\ket{\Psi}^{-}$ and $\ket{\Phi}^{+}$ with the help of polarization-dependent phase shift (phase-flip) and polarization exchange (bit-flip) respectively. Conventionally, A quarter-wave plate (QWP) (which performs a Pauli-Z transformation) and half-wave plate (HWP) (which performs a Pauli-X transformation) is used for phase-flip and bit-flip respectively \cite{kwiat1995new,kong2015near, wang2017generation}. In this work, we show that only a single half-wave plate is enough to perform Pauli-X and Pauli-Z transformation. A pair of HWP would be required to perform Pauli-Y transformation. \par

This paper is organized as follows: In section \ref{sec:theory}, we explain the theoretical model of our proposed scheme where we first explain the single qubit gate (Pauli gate) and how these gates are analogous to the rotation of HWP.
In section \ref{sec:exp} we present the experimental setup for both generation and transformation of Bell states. In section \ref{sec:res}, we discuss the experimental results. We end the paper with concluding remarks in last section \ref{sec:conc}.

\section{\label{sec:theory}Theory}
\begin{figure}[h]
    \centering
   \includegraphics[width=8.5cm]{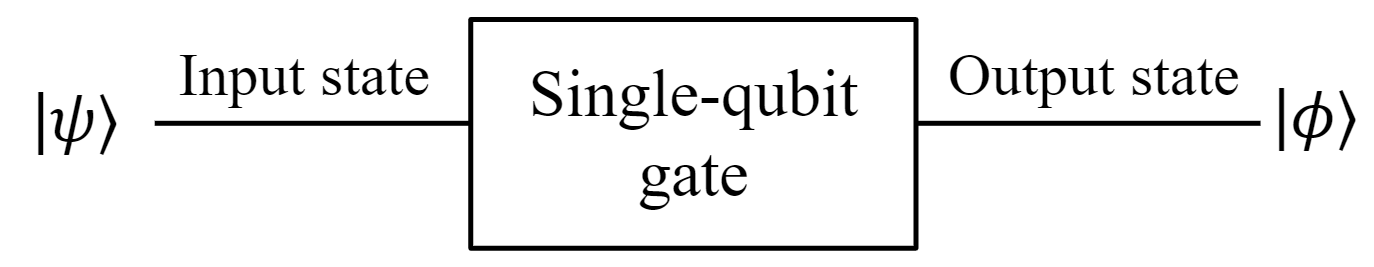}
   \caption{Schematic diagram of transformation of state}
   \label{sch}
   \end{figure}

\begin{figure}[h]
\includegraphics[width=8.5cm]{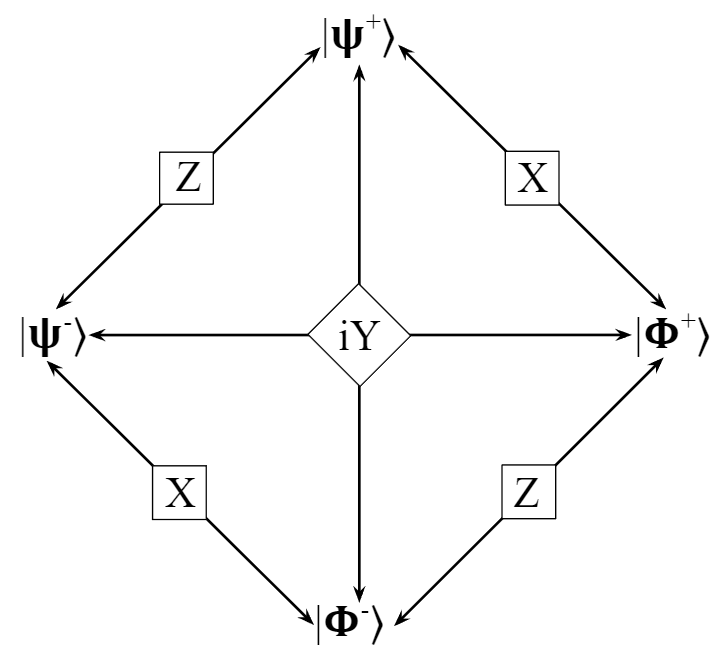}
\caption{Transformation of Bell states using single-qubit logic gates}
\label{fig:sqlg}
\end{figure}

\noindent Consider two photons, signal and idler, are entangled in polarization and propagating along two different directions. This entangled state can be expressed as in four different ways,
\begin{subequations}
\begin{equation}
\ket{\Psi}^{\pm}=\frac{1}{\sqrt{2}}(\ket{H}_{s}\ket{V}_{i}\pm\ket{V}_{s}\ket{H}_{i}),
\end{equation}
\begin{equation}
\ket{\Phi}^{\pm}=\frac{1}{\sqrt{2}}(\ket{H}_{s}\ket{H}_{i}\pm\ket{V}_{s}\ket{V}_{i}).
\end{equation}
\label{eq1}
\end{subequations}
where $\ket{\Psi}^{-}$ is anti-symmetric entangled state and rest are symmetric states.These polarization entangled states are popularly known as Bell states. $\ket{H}$ and $\ket{V}$ are the horizontal and vertical polarization state of photons, respectively. Using Jones vector notation, the polarization states can be represented by column vector,
\begin{subequations}
\begin{equation}
\ket{H}=\left(
\begin{array}{cc}
1\\
0\\
\end{array}
\right),\,\,
\ket{V}=\left(
\begin{array}{cc}
0\\
1\\
\end{array}
\right)
\end{equation}
\begin{equation}
\ket{D}=\left(
\begin{array}{cc}
1\\
1\\
\end{array}
\right), \,\,
\ket{A}=\left(
\begin{array}{cc}
1\\
-1\\
\end{array}
\right)
\end{equation}
\label{eq2}
\end{subequations}
Using single-qubit quantum logic gate in Eqn. \ref{eq1}, one Bell state can be easily manipulated and transformed into another Bell state. Such a gate operation transforms input state (say $\ket{\psi}$) to an output state   (say $\ket{\phi}$) ( Figure \ref{sch}). 
These single-qubit logic gates are Paulie $X$, $Y$, and $Z$ gate and are analogous to Pauli spin matrices $\sigma_{x}$, $\sigma_{y}$ and $\sigma_{z}$ respectively. The matrix representation for these gates is given as,
\begin{equation}
\begin{aligned}
I&=\sigma_{0}=\left(
    \begin{array}{cc}
1&0\\
0&1\\
\end{array}
\right),\,\,
\hspace{5pt}X=\sigma_{x}=\left(
    \begin{array}{cc}
0&1\\
1&0\\
\end{array}
\right)\\
Y&=\sigma_{y}=\left(
    \begin{array}{cc}
0&-i\\
i&0\\
\end{array}
\right),\,\,
Z=\sigma_{z}=\left(
    \begin{array}{cc}
1&0\\
0&-1\\
\end{array}
\right)
\end{aligned}
\label{eq3}
\end{equation}
Here, $I$ is the identity gate and it does not change the input state. $X$ gate is known as bit-flip because it flips the state $\ket{H}$ to $\ket{V}$ and vice-versa, whereas $Z$ gate is known as  phase-flip because it flips the sign of $\ket{V}$, while leaving $\ket{H}$ unchanged. $iY$ gate is used to flip the bit and phase simultaneously and it can be realized using the property $[\sigma_{x},\,\sigma_{z}]=2i\sigma_{y}$. Therefore, 
\begin{equation}
\begin{aligned}
\sigma_{x}\sigma_{z}&=i\sigma_{y}\\
 XZ&=iY=\left(
    \begin{array}{cc}
0&1\\
-1&0\\
\end{array}
\right) 
\end{aligned}
\label{eq3.0}
\end{equation}
Now consider our initial state is $\ket{\Psi}^{+}$ (Eqn. \ref{eq1}). 
Applying $Z$ and $X$ gate operation on signal photon of state $\ket{\Psi}^{+}$ results $\ket{\Psi}^{-}$ and $\ket{\Phi}^{+}$ respectively,
\begin{subequations}
\begin{align}
    Z_{s}\ket{\Psi}^{+}&=\ket{\Psi}^{-}\\
    X_{s}\ket{\Psi}^{+}&=\ket{\Phi}^{+}
\end{align}
Again, applying $Y$ gate operation on signal photon results $\ket{\Phi}^{+}$ and $\ket{\Phi}^{-}$ respectively,
\begin{align}
iY\ket{\Psi}^{+}&=X_{s}Z_{s}\ket{\Psi}^{+}=\ket{\Phi}^{-}
\end{align}
Or one can apply $X$ gate in signal photon and $Z$ gate on idler photon and vice-versa. Hence $Y$ gate operation can also be written as,
\begin{equation}
   iY\ket{\Psi}^{+}= X_{s}Z_{i}\ket{\Psi}^{+}=\ket{\Phi}^{-}
\end{equation}
\label{eq4}
\end{subequations}
These states transformation are illustrated in Fig. \ref{fig:sqlg}.
These gates can be experimentally realized in laboratory using a half-wave plate (HWP). HWP is used to rotate the polarization of photons \cite{ratajczyk2000transformation,reddy2016polarization}. It retards the polarization state of photons by adding a phase difference between the two orthogonal polarization components $\ket{H}$ and $\ket{V}$ and the phase difference depends on wavelength $\lambda$ of incident photons, birefringence properties ($\Delta n$) of the crystal, and thickness $d$ of the HWP,
\begin{equation}
\Delta\phi=\frac{2\pi d \Delta n}{\lambda}
\label{eq5}
\end{equation}
where $\Delta n$ is the difference of refractive index along slow axis and fast axis. $\lambda$, $\Delta n$, and $d$ are chosen in such a way that the phase difference between polarization components is $\pi$. The action of HWP can be represented by Jones matrix notation,
\begin{equation}
 \hat{U}_{\text{hwp}}(\theta)= \left(
\begin{array}{cc}
\text{cos}\, 2\theta&\text{sin}\, 2\theta\\
\text{sin}\, 2\theta&-\text{cos}\, 2\theta\\
\end{array}
\right)  
\label{eq6}
\end{equation}
where $\theta$ is the angle between fast axis and horizontal axis.
When the photons of particular polarization transmit through the HWP, the polarization changes according to HWP angle $\theta$. $Z$ and $X$ gate can be realized by aligning the fast axis of HWP at an angle 0$^{\circ}$ and 45$^{\circ}$ with respect to horizontal axis and $i$Y gate can be realized by using two HWPs at two different angles 0$^{\circ}$ and 45$^{\circ}$.
\begin{equation}
\begin{aligned}
 \hat{U}_{\text{hwp}}(0^{\circ})&=\left(
\begin{array}{cc}
1&0\\
0&-1\\
\end{array}
\right)=\sigma_{z}\\
\hat{U}_{\text{hwp}}(45^{\circ})&=\left(
\begin{array}{cc}
0&1\\
1&0\\
\end{array}
\right)=\sigma_{x}\\
\hat{U}_{\text{hwp}}(0^{\circ}) \hat{U}_{\text{hwp}}(45^{\circ})&=\left(
\begin{array}{cc}
0&1\\
-1&0\\
\end{array}
\right) = i\sigma_{y}
\end{aligned}
\label{eq7}
\end{equation}
Using Eqn. \ref{eq2} and \ref{eq7}, we will show that how one can use only HWP to perform all quantum gates,\\

\vspace{5pt}
\noindent I. Verification of Z gate:
\begin{subequations}
\begin{align}
\hat{U}_{\text{hwp}}(0^{\circ})\ket{H}&=\;\ket{H},\hspace{10pt}
\hat{U}_{\text{hwp}}(0^{\circ})\ket{V}=-\ket{V},\\
\hat{U}_{\text{hwp}}(0^{\circ})\ket{D}&=\;\ket{A},\hspace{10pt}
\hat{U}_{\text{hwp}}(0^{\circ})\ket{A}=\;\ket{D}.
\end{align}
\vspace{5pt}
\label{eq8}
\end{subequations}
II. Verification of X gate:
\begin{subequations}
\begin{align}
\hat{U}_{\text{hwp}}(45^{\circ})\ket{H}&=\ket{V},\hspace{10pt}
\hat{U}_{\text{hwp}}(45^{\circ})\ket{V}=\ket{H}\\
\hat{U}_{\text{hwp}}(45^{\circ})\ket{D}&=\ket{D},\hspace{10pt}
\hat{U}_{\text{hwp}}(45^{\circ})\ket{A}=-\ket{A}
\end{align}
\label{eq9}
\end{subequations}
Eqn. \ref{eq8} shows that apart from quarter-wave plate (QWP),  HWP can also be used for phase-flip.
Using Eqn. \ref{eq8} and \ref{eq9}, one can transform bell states using HWPs only.
Table \ref{bellstate} shows the transformation of Bell state using HWP.
\begin{table}[h]

    \begin{tabular}{m{1.3cm}|m{1.7cm}|m{1.3cm}|m{1.3cm}|m{1.3cm}}
        \hline
        \vspace{2pt}
       \textbf{Input state} & \textbf{Quantum gate} &\hspace{2pt} \textbf{HWP$_{a}$} &\hspace{2pt} \textbf{HWP$_{b}$} & \textbf{Output state} \\ 
        \hline 
       \hspace{6pt} $\ket{\Psi}^{+}$ &\hspace{16pt} I  &  &  &\hspace{6pt}$\ket{\Psi}^{+}$ \\
        \hline
       \hspace{6pt} $\ket{\Psi}^{+}$ & \hspace{16pt} Z  &\hspace{15pt} 0$^{\circ}$ &  &\hspace{6pt}$\ket{\Psi}^{-}$ \\ 
        \hline 
      \hspace{6pt}  $\ket{\Psi}^{+}$ &\hspace{15pt} X  &\hspace{15pt} 45$^{\circ}$ &  &\hspace{6pt}$\ket{\Phi}^{+}$ \\ 
        \hline 
     \hspace{6pt} $\ket{\Psi}^{+}$ &\hspace{14pt} XZ  &  \hspace{15pt} 0$^{\circ}$ & \hspace{15pt} 45$^{\circ}$ &\hspace{6pt}$\ket{\Phi}^{-}$ \\
     \hline
    \end{tabular} 
   \caption{Operation of quantum gate and corresponding HWP angles. The gate transforms an input state to an output state.}
\label{bellstate}
\end{table}

\section{\label{sec:exp}Experimental setup}
\subsection{\label{sec:epr}Entangled photon pair generation}
\begin{figure}[h]
\centering
\includegraphics[width=9cm]{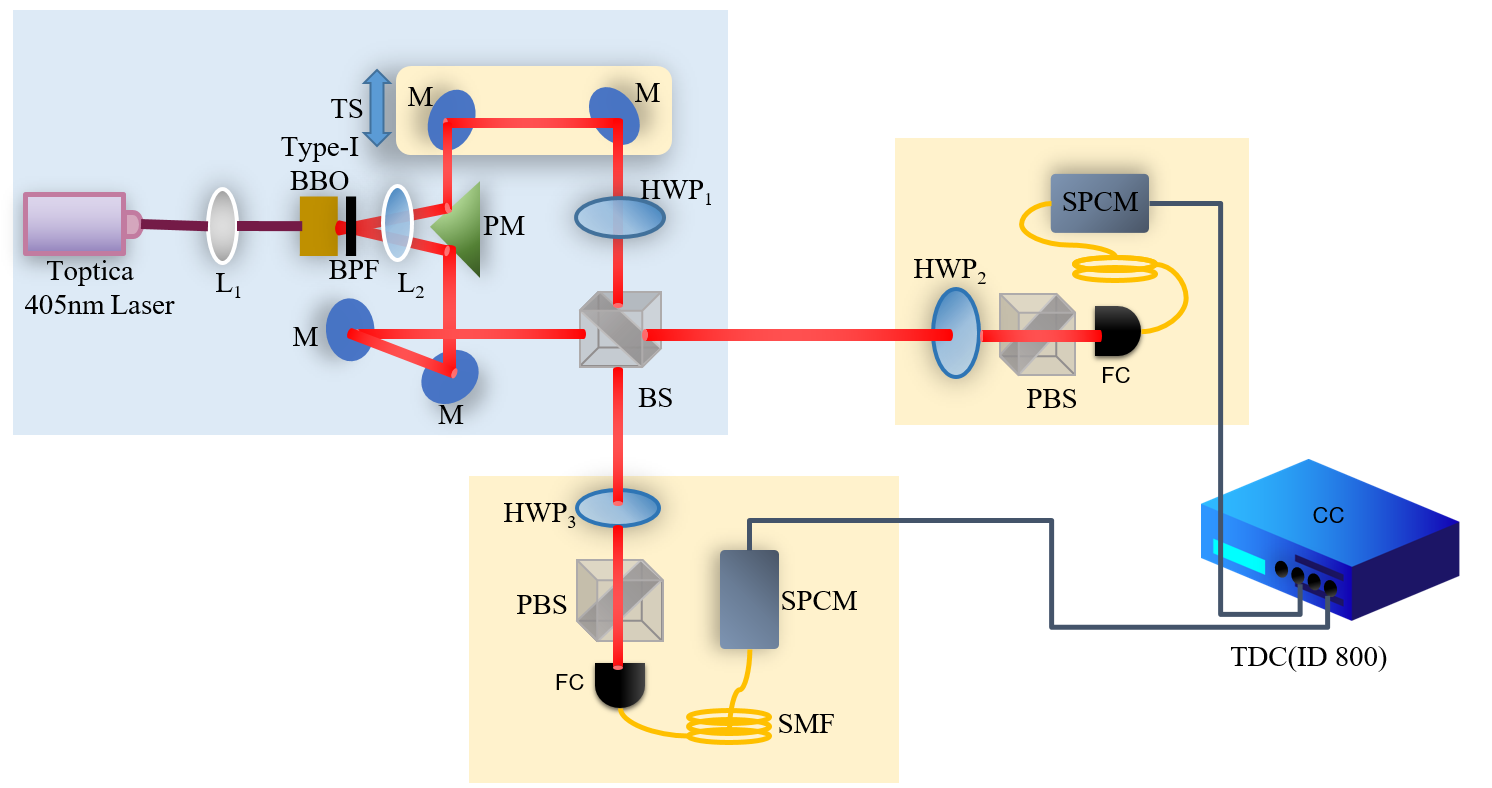}
\caption{Experimental setup for the generation of entangled state $\ket{\Psi}^{+}$. L$_{1}$, L$_{2}$ - Lenses, BBO -  $\beta$-Barium borate crystal, BPF - Band pass filter, PM - Prism mirror, M - Mirror, BS - 50:50 Beam splitter, PBS - Polarizing beam splitter, HWP - Half-wave plate, FC - Fiber coupler, SMF - Single mode fiber, SPCM - single photon counting module, CC - Coincidence counter }
\label{fig:HOM}
\end{figure}

\noindent Fig. \ref{fig:HOM} shows the experimental setup for the generation of polarization entangled Bell state $\ket{\Psi}^{+}$. Horizontally polarized continuous wave laser (Toptica iBeam smart) of wavelength 405 nm and pump power 2.5 mW is used to pump the nonlinear type-I $\beta$-barium borate ($\beta$-BBO) crystal of thickness 2 mm and transverse dimensions of 6 mm × 6 mm with an optic axis oriented at $29.97^{\circ}$ to the normal incidence. This will produce two vertically polarized photon pair of wavelength 810 nm each. A band pas filter (BPF) is used to block 405 nm photons while transmitting the 810 nm photons. Both the photons are separated by prism mirror (PM) in two different arms. Both the photons are again combined at the input port of the 50:50 beam splitter (BS). A half-wave plate (HWP$_{1}$) is used in one of the input arm of beam-splitter (BS) to flip the polarization state of one photon so that both the photons become orthogonally polarized to each other. A translation delay stage is used to compensate the path difference between signal and idler. \par
For the detection part, we have used combination of HWP and PBS at each output port of BS. After combining at the BS both the photons will exit from two different output ports and get coupled into single mode fiber (SMF) (P1-780A-FC-2, Thorlabs) with the help of fiber couplers (FC) (CFC-5X-B, Thorlabs) . These SMFs are connected to the single photon counting module (SPCM) detectors (SPCM-AQRH-16-FC, Excelitas) having a time resolution 350 ps. To measure the number of correlated photon pairs, Both the SPCMs are then connected to a coincidence counter (CC), IDQuantique-ID800, having a time resolution of 81 ps. The photons pairs that exit from same output port will not contribute to coincidence counts.
This setup can directly produce $\ket{\Psi}^{\pm}$ state at the output ports of BS by simply adjusting the phase difference between both the photons.

\begin{figure}[h!]
\centering
\includegraphics[width=9cm]{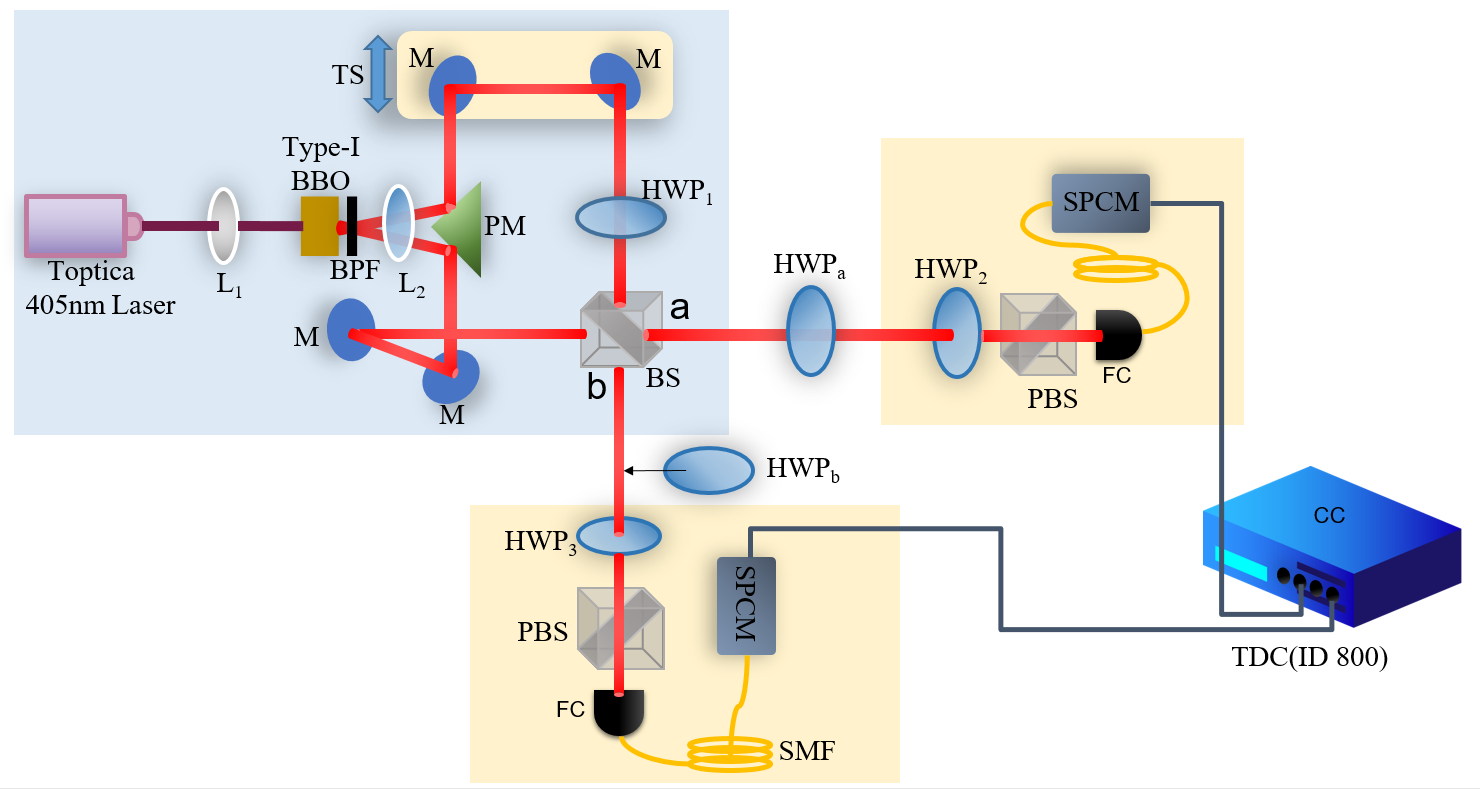}
\caption{Experimental setup for the transformation of entangled state $\ket{\Psi^{+}}$ into $\ket{\Psi}^{-}\,\text{and}\, \ket{\Phi}^{+} $ by performing $Z$ and $X$ gate operation respectively using HWP$_a$ only. Additional HWP$_b$ along with HWP$_a$ is used to transform the state $\ket{\Psi^{+}}$ into $\ket{\Phi}^{-}$.}
\label{fig:xzgate}
\end{figure}
\subsection{\label{sec:trans}Transformation of Bell state}
\noindent Fig. \ref{fig:xzgate} shows the experimental setup of the transformation of Bell state. First we generated Bell state $\ket{\Psi^{+}}$ using Fig. \ref{fig:HOM} and then in the same setup, we kept HWP$_a$ and HWP$_{b}$ in the output port of BS as shown in Fig. \ref{fig:xzgate}. We transform state $\ket{\Psi^{+}}$ into states $\ket{\Psi}^{-}$ and $\ket{\Phi}^{+}$ by performing $Z$ and $X$ gate respectively in output port $a$ using HWP$_{a}$. One extra HWP$_{b}$ along with  HWP$_{a}$ is used to produce $\ket{\Phi}^{-}$ state by performing $Y$ gate operation. In this case both the HWPs, a and b, are rotated at different orientations 0$^{\circ}$ and 45$^{\circ}$ respectively. In our case, both HWP is placed at output port $a$ and $b$, but one can put both HWPs at same output port. Both the cases will give the same result.

\section{\label{sec:res}Results and discussion}
\noindent In this section, we present experimental results obtained for all the four polarization-entangled Bell states. In Fig. \ref{fig:xzgate}a, when HWP$_{a}$ is aligned to 0$^{\circ}$  then it will act as Z gate and hence it will flip the sign of $\ket{V}_{a}$, while leaving $\ket{H}_{a}$ unchanged. In D/A basis, $\ket{D}_{a}$ will convert to $\ket{A}_{a}$ and vice-versa. This will transferred the state from $\ket{\Psi}^{+}$ to $\ket{\Psi}^{-}$.
The initial state is given by,
\begin{equation}
\ket{\Psi}^{+} =
\left\{
\begin{array}{lr}
\frac{1}{\sqrt{2}}(\ket{H}_{a}\ket{V}_{b}+\ket{V}_{a}\ket{H}_{b}) & \,H/V\, \text{basis}\\
\frac{1}{\sqrt{2}}(\ket{D}_{a}\ket{D}_{b}-\ket{A}_{a}\ket{A}_{b}) & \,D/A\, \text{basis}
\end{array}
\right.
\end{equation}
\vspace{5pt}\\
For $\theta_{a}=0^{\circ}$, output state will be,
$$U_{\text{hwp}}(\theta_{a}=0^{\circ})\ket{\Psi}^{+}=\ket{\Psi}^{-}$$
\begin{equation}
\ket{\Psi}^{-} =
\left\{
\begin{array}{lr}
\frac{1}{\sqrt{2}}(\ket{H}_{a}\ket{V}_{b}-\ket{V}_{a}\ket{H}_{b}) & \,H/V\, \text{basis}\\
\frac{1}{\sqrt{2}}(\ket{A}_{a}\ket{D}_{b}-\ket{D}_{a}\ket{A}_{b}) & \,D/A\, \text{basis}
\end{array}
\right.
\end{equation}
Now if we again rotate the HWP$_{a}$ to 45$^{\circ}$ then it will act as a X gate and hence it will convert $\ket{H}_{a}$ to $\ket{V}_{a}$ and vice-versa. Also it will flip the sign of $\ket{A}_{a}$, while leaving $\ket{D}_{a}$ unchanged.Therefore, for $\theta_{a}=45^{\circ}$, the state will become,
For $\theta_{a}=45^{\circ}$,
$$U_{\text{hwp}}(\theta_{a}=45^{\circ})\ket{\Psi}^{+}=\ket{\Phi}^{+}$$
\begin{equation}
\ket{\Phi}^{+} =
\left\{
\begin{array}{lr}
\frac{1}{\sqrt{2}}(\ket{H}_{a}\ket{H}_{b}+\ket{V}_{a}\ket{V}_{b}) & \,H/V\, \text{basis}\\
\frac{1}{\sqrt{2}}(\ket{D}_{a}\ket{D}_{b}+\ket{A}_{a}\ket{A}_{b}) & \,D/A\, \text{basis}
\end{array}
\right.
\end{equation}
Now to transfer $\ket{\Psi}^{+}$ state to $\ket{\Phi}^{-}$ state we used one more HWP$_{b}$ in output port b of BS (Fig. \ref{fig:xzgate}b). we aligned HWP$_{a}$ to 0$^{\circ}$  and HWP$_{b}$ to 45$^{\circ}$ to get $\ket{\Phi}^{-}$ state. Therefore, for $\theta_{a}=0^{\circ}$ and $\theta_{b}=45^{\circ}$, the output state will be,
$$U_{\text{hwp}}(\theta_{a}=0^{\circ})U_{\text{hwp}}(\theta_{b}=45^{\circ})\ket{\Psi}^{+}=\ket{\Phi}^{-}$$
\begin{equation}
\ket{\Phi}^{-} =
\left\{
\begin{array}{lr}
\frac{1}{\sqrt{2}}(\ket{H}_{a}\ket{H}_{b}-\ket{V}_{a}\ket{V}_{b}) &  \,H/V\, \text{basis}\\
\frac{1}{\sqrt{2}}(\ket{A}_{a}\ket{D}_{b}+\ket{D}_{a}\ket{A}_{b}) &  \,D/A\, \text{basis}
\end{array}
\right.
\end{equation}
\begin{figure}[h!]
    \centering
  \fbox{(a) \includegraphics[width=7.5cm]{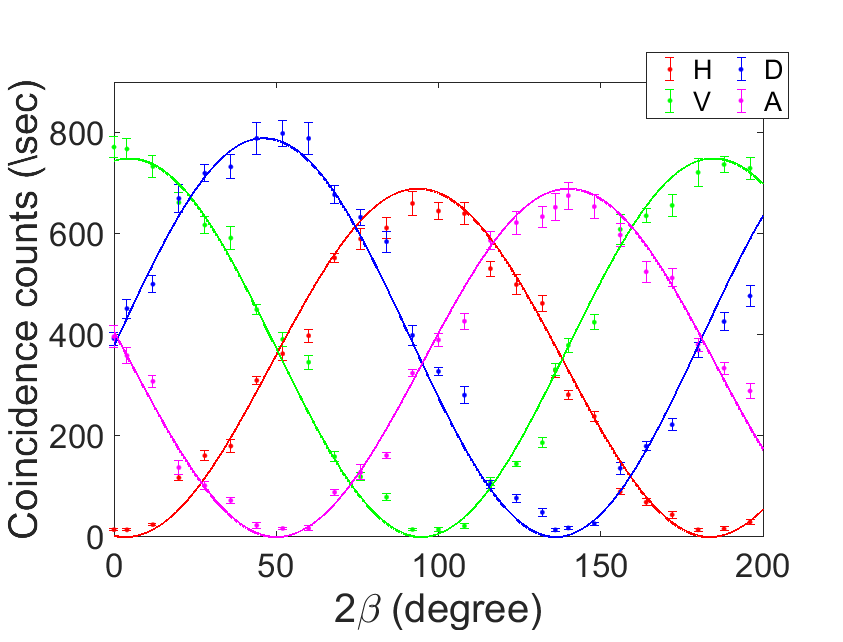}}
  \fbox{(b) \includegraphics[width=7.5cm]{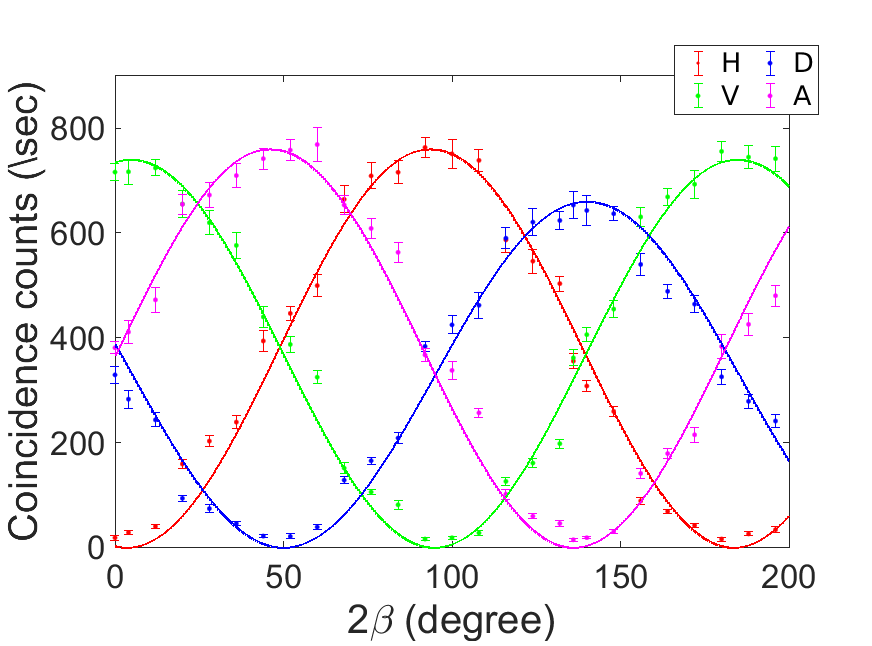}}
  \caption{Experimental observation of polarization correlations for (a) $\ket{\Psi}^{+}$ and (b) $\ket{\Psi}^{-}$ state. Error bars
indicate statistical uncertainty of one standard deviation.}
    \label{fig:psi}
\end{figure}
We have recorded the measurement in coincidence basis. Polarization measurements are carried out using a combination of HWP and PBS in each arm. This combination will act as a projection operator. The measurement is performed by fixing the angle of HWP$_{2}$, $\alpha$ in one output port, while changing the angle of HWP$_{3}$, $\beta$ in another port. The transmitted part of PBS is coupled into a single mode fiber. For example, if we want to measure the diagonal polarization (D) of photon then first we will rotate the diagonal polarization (D) to horizontal (H) with the help of HWP so that photons get transmitted through the PBS and get coupled into single mode fiber.\par

\begin{table}[h!]
    \centering
    \begin{tabular}{m{2cm}|m{1.2cm}|m{1.2cm}|m{1.2cm}|m{1.2cm}}
        \hline
        \vspace{2pt}
        \textbf{Coincidence detection} & $\ket{\Psi}^{+}$ & $\ket{\Psi}^{-}$ & $\ket{\Phi}^{+}$ &$\ket{\Phi}^{-}$  \\ 
        \hline
        \vspace{2pt}
        \textbf{HH/VV} & -  & - & \checkmark & \checkmark \\ 
        \hline 
        \vspace{2pt}
       \textbf{HV/VH}  & \checkmark  & \checkmark & -  & - \\ 
        \hline 
        \vspace{2pt}
       \textbf{DD/AA}  & \checkmark  & - & \checkmark  & - \\ 
        \hline 
        \vspace{2pt}
       \textbf{DA/AD}  & -  & \checkmark & - & \checkmark \\ \hline
    \end{tabular} 
    \caption{Theoretical prediction of Detecting entangled photon pair in H/V/D/A basis}
    \label{theory}
\end{table}
\begin{table}[h!]
    \centering
    \begin{tabular}{m{2cm}|m{1.4cm}|m{1.4cm}|m{1.4cm}|m{1.4cm}}
        \hline \vspace{2pt}
        \textbf{Coincidence detection} & $\ket{\Psi}^{+}$ & $\ket{\Psi}^{-}$ & $\ket{\Phi}^{+}$ &$\ket{\Phi}^{-}$  \\ 
        \hline \vspace{2pt}
       \textbf{HH/VV} & 13/14   & 18/15   & 698/751 & 731/784 \\ 
        \hline \vspace{2pt}
       \textbf{HV/VH} & 660/771 & 763/724 & 21/14   & 18/15 \\ 
        \hline \vspace{2pt}
       \textbf{DD/AA} & 798/675 & 17/14   & 680/738 & 17/15 \\ 
        \hline \vspace{2pt}
       \textbf{DA/AD} & 13/16   & 653/768 & 14/15  & 825/735 \\ \hline 
    \end{tabular} 
    \caption{Experimental observation of detecting counts of entangled photon pair in H/V/D/A basis. The coincidence counts are recorded with a coincidence window of 1.62 ns and integration time of 1 s.}
    \label{exp}
\end{table}
\begin{figure}[h!]
    \centering
  \fbox{(a) \includegraphics[width=7.5cm]{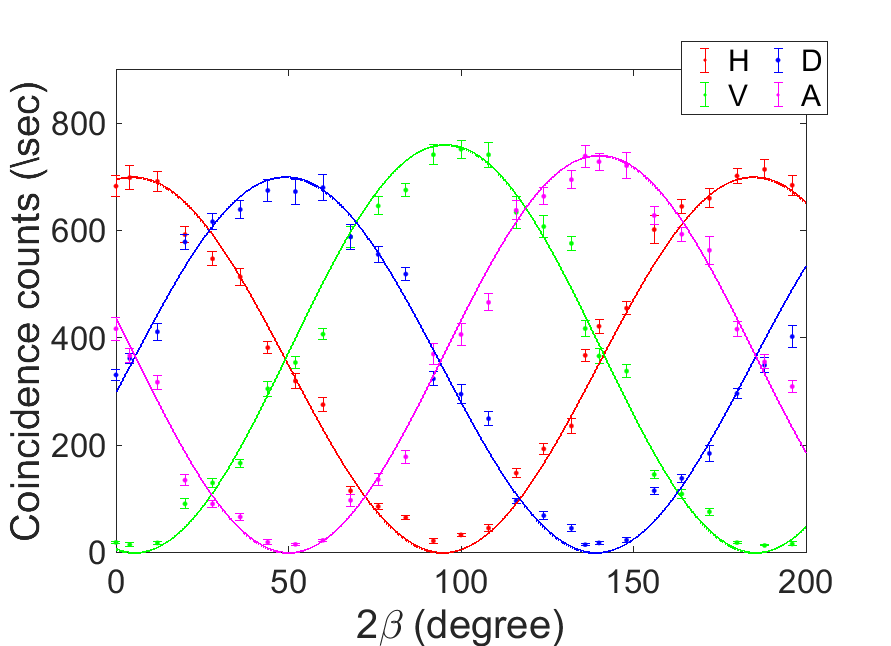}}
  \fbox{(b) \includegraphics[width=7.5cm]{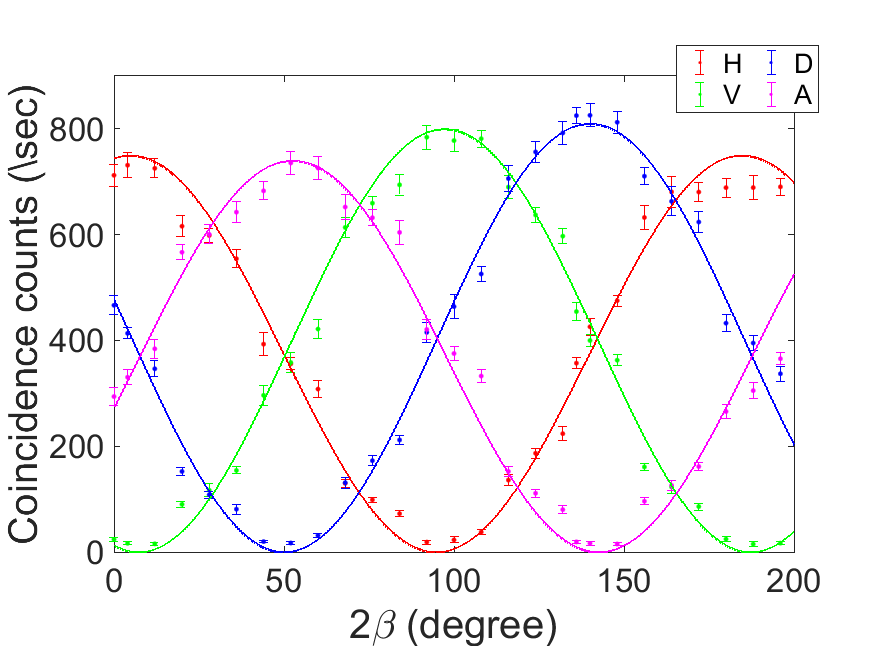}}
  \caption{Experimental observation of polarization correlations for (a) $\ket{\Phi}^{+}$ and (b) $\ket{\Phi}^{-}$ state. Error bars
indicate statistical uncertainty of one standard deviation.}
    \label{fig:phi}
\end{figure}
Table \ref{theory} shows the theoretical discrimination between all the four Bell state. To verify the theoretical prediction, we have recorded coincidence counts per second in H/V and D/A basis for each Bell state. Coincidence detection is experimentally measured number of photon pairs for certain polarization state. Table \ref{exp} shows the experimental data. Our experimental result is in excellent agreement  with the theoretical prediction and by comparing table \ref{theory} and \ref{exp}, one can predict which bell state is measured. The experimentally observed polarization correlations are given as visibility curves in Fig. \ref{fig:psi} and \ref{fig:phi}. The total coincidence counts per second are plotted in y axis as a function of HWP$_{3}$ angle $2\beta$, while HWP$_{2}$ angle $2\alpha$ is fixed. \par
 To check the quality of entanglement, we calculated the Bell-CHSH parameter S. The S value greater than 2 ensures the existence of quantum entanglement between two photons. In our experiment, the average value of visibilities and the estimated Bell-CHSH parameter S for $\ket{\Psi}^{+}$, $\ket{\Psi}^{-}$,$\ket{\Phi}^{+}$, and $\ket{\Phi}^{-}$ are shown in table \ref{Svalue}. Because of the unitary transformation the S value for all the four Bell state is remains unchanged.  
\begin{figure}[h!]
    \centering
    \includegraphics[width=8.5cm]{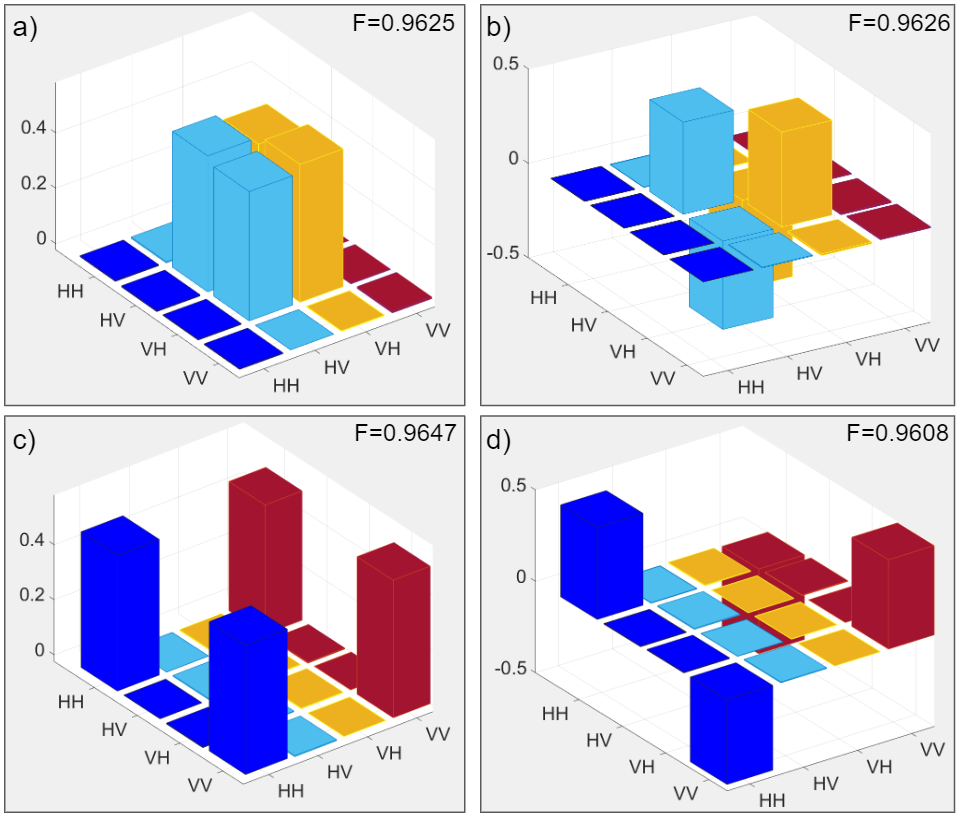}
    \caption{Density matrices are calculated for entangled states (a)$\ket{\Psi}^{+}$ and (b) $\ket{\Psi}^{-}$ (c)$\ket{\Phi}^{+}$ and (d) $\ket{\Phi}^{-}$. F is the state fidelity which is the measure of state overlap }
    \label{tomo}
\end{figure}
\begin{table}[h!]
\centering
\begin{tabular}{l | l | l|l}
 \hline 
\textbf{Bell state} & \textbf{Avg. Vis.($\%$)} & \textbf{S Value}& \textbf{Fidelity (F)} \\
\hline 
$\ket{\Psi}^{+}$ & 96.18  $\pm$ 5.03  & 2.72 $\pm$ 0.07 &\hspace{5pt} 0.9625\\
\hline 
$\ket{\Psi}^{-}$ & 95.66$\pm$ 4.28  & 2.71 $\pm$ 0.06 &\hspace{5pt} 0.9626\\
\hline 
$\ket{\Phi}^{+}$ & 95.61 $\pm$ 3.75 & 2.70 $\pm$ 0.05 & \hspace{5pt} 0.9647\\
\hline 
$\ket{\Phi}^{-}$ & 95.84 $\pm$ 3.17 & 2.71 $\pm$ 0.05 & \hspace{5pt} 0.9608\\
\hline 
\end{tabular}
\caption{Calculated average visibility,Bell-CHSH parameter S, and fidelity F.}
\label{Svalue}
\end{table}
Quantum state tomography is performed  to verify the theoretical prediction by calculating the state fidelity. The density matrix for all the four Bell states are shown in Fig.\ref{tomo}. The state fidelity is around  $\sim96\%$ for each Bell state which shows the successful transformation of Bell states using HWPs. 

\section{\label{sec:conc}Conclusion}
\noindent In this article, we have shown that only half-wave plate (HWP) is enough to perform Pauli Z gate and X gate and it become the easiest way to transform entangled state. We first prepared the polarization-entangled Bell state $\ket{\Psi}^{+}$ using type-I $\beta$-BBO crystal and 50:50 beam-splitter (BS) and then manipulated the initial state in order to achieve rest of the entangled state. We calculated the Bell-CHSH parameter S for each Bell state and since the transformation is unitary, we found out that S value, coincidence counts, and visibility for all the Bell state remains unchanged. Quantum state tomography is also performed to check the state fidelity with the desired state. The presented result may find applications in quantum communication and quantum information protocols, especially where the control over basis of maximally entangled state is required.

\section*{Disclosures}
\noindent The authors declare no conflicts of interest related to this article

%

\end{document}